\documentclass[11pt,twoside]{article}

\usepackage{asp2006}
\usepackage{epsf}
\usepackage{psfig}

\begin{document}
\title{H$\alpha$ kinematics of KPG 390.}   
\author{P. Repetto, M. Rosado, R. Gabbasov}   
\affil{Instituto de Astronom\'{\i}a, Universidad Nacional
Autonoma de M\'exico (UNAM), Apdo. Postal 70-264, 04510,
M\'exico, D.F., M\'exico}

\author{I. Fuentes-Carrera}
\affil{Department of Physics, Escuela
Superior de F{\'i}sica y Matem\'aticas, IPN, U.P. Adolfo L\'opez Mateos,
C.P. 07738, Mexico city, M\'exico.}    

\begin{abstract} 
In this work we present scanning Fabry-Perot
H$\alpha$ observations of the isolated interacting
galaxy pair NGC 5278/79 obtained with the PUMA Fabry-Perot interferometer.
We derived velocity fields, various kinematic parameters
and rotation curves for both galaxies.
Our kinematical results together with the fact that dust lanes
have been detected in both galaxies, as well as the analysis of surface
brightness profiles along the minor axis, allowed us to determine
that both components of the interacting pair are trailing spirals.
\end{abstract}

\section{Introduction}   

Interactions and mergers of galaxies are common
phenomena in the Universe. Isolated pairs of galaxies
represent a relatively easy way to study
interactions between galaxies because these systems,
from a kinematical point of view, are simpler than
associations and compact groups of galaxies, where
so many galaxies participate in the interaction process,
that it is difficult to discriminate the role of each
galaxy in the interaction. In this work we present
scanning Fabry-Perot observations, velocity fields
and rotation curves of this interacting galaxy
pair. The aim of this study is to perform detailed kinematic
and dynamic analysis of NGC 5278/79 (Arp 239 KPG 390)
using H$\alpha$ kinematical data in order to study
the mass distribution of this pair of galaxies 
and to determine the type of spiral arms
(leading or trailing) in
the galaxy members with the intention, in a future work, 
of reproduce both its morphology and
kinematics with numerical simulations that could  shed more light
on the interaction process.

\section{Velocity fields}

\begin{figure}[!htp]
\epsscale{0.3}\plottwo{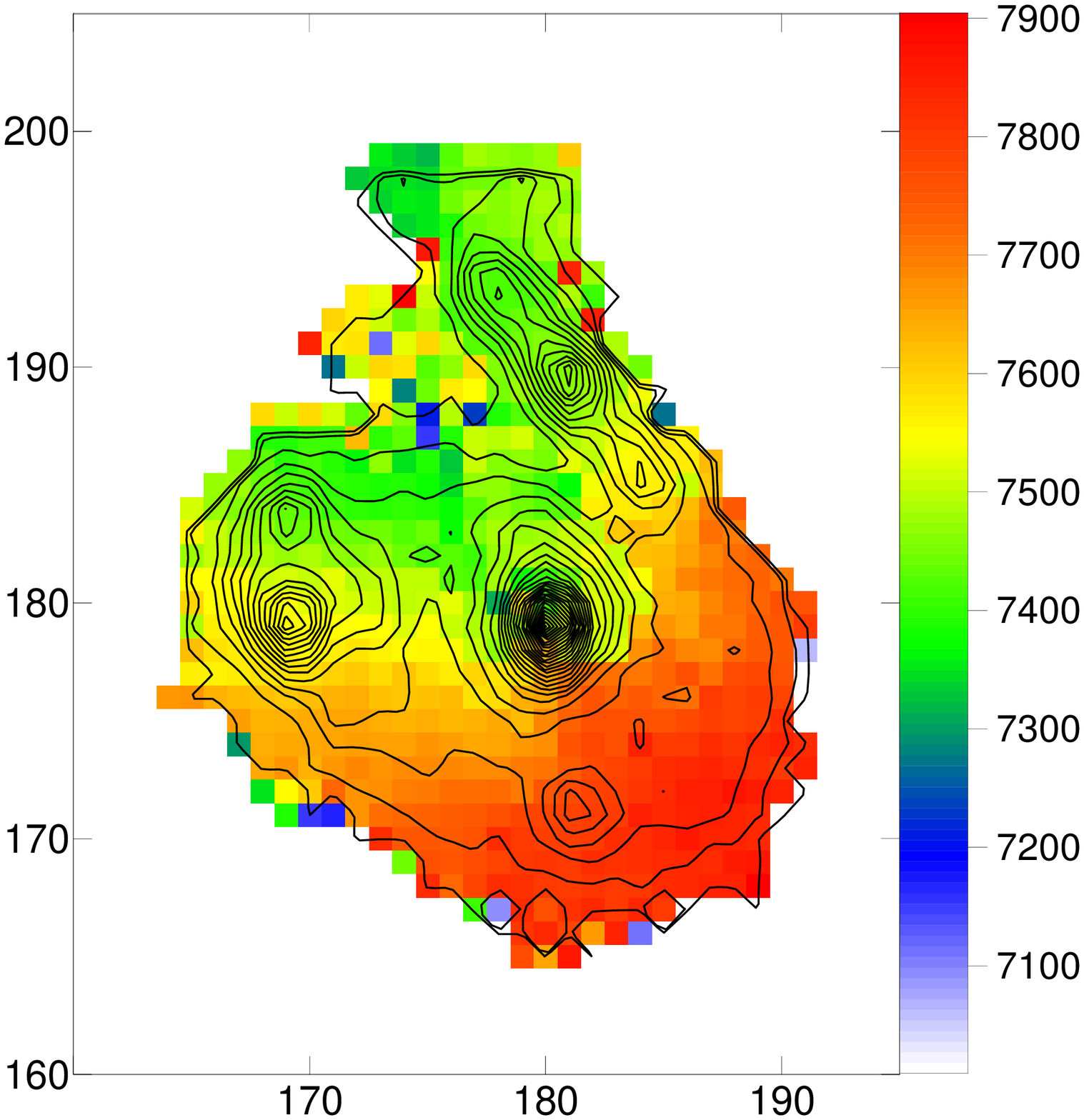}{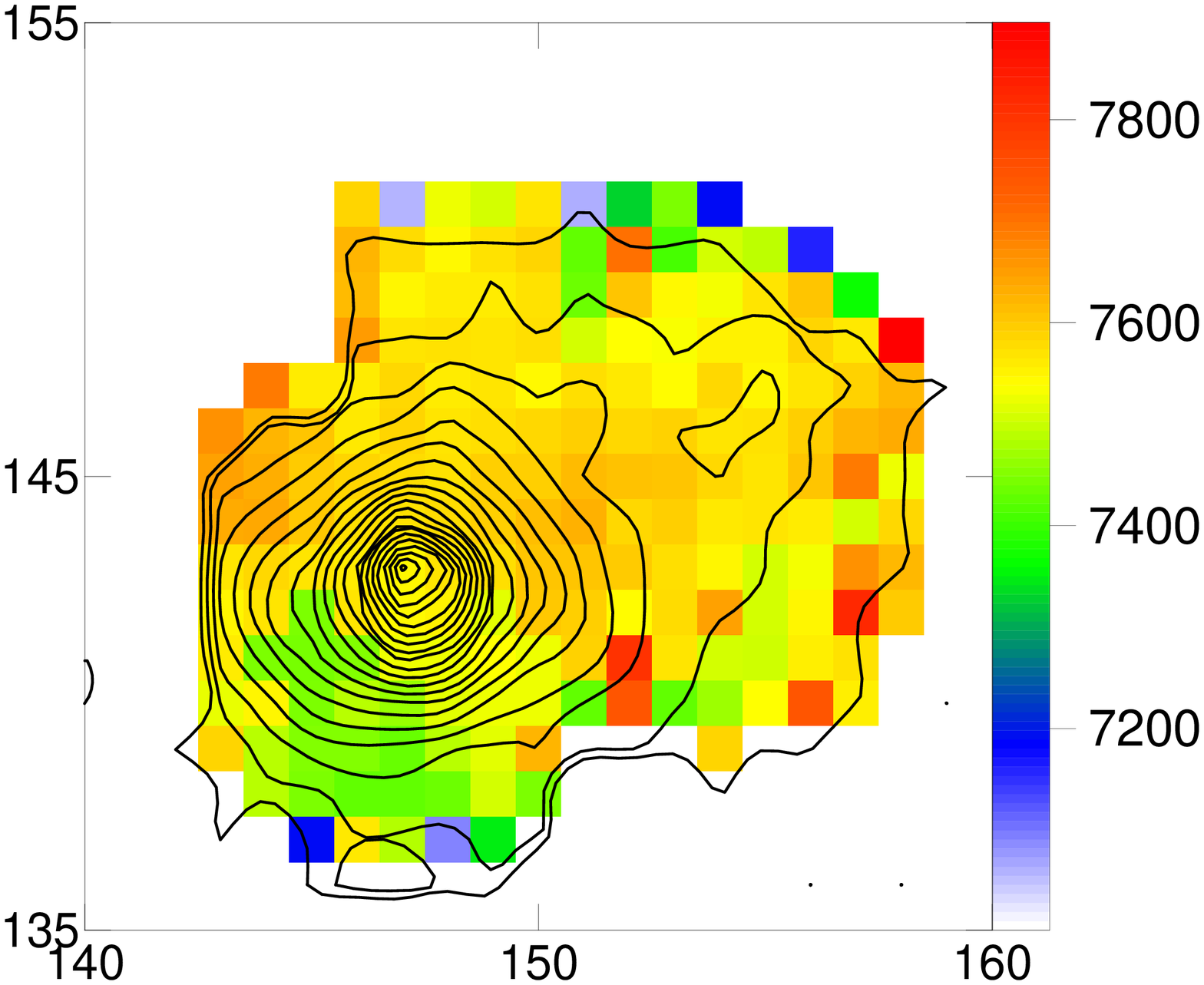} 
\caption{{\it Left}: Velocity field of NGC 5278 obtained from the FP
velocity cubes with the H$\alpha$ image isophotes superposed. 
{\it Right}: Velocity field of NGC 5279 obtained from the FP
velocity cubes with the monochromatic map isophotes superposed.
The numbers in the color scales are heliocentric systemic velocities in 
km s$^{-1}$.\label{fig1}}
\end{figure}

The velocity fields of NGC 5278 (primary) and NGC 5279
(secondary)
are shown in Fig.~\ref{fig1}. The velocity field 
of NGC 5278 shows a rather smooth
behaviour similar to the velocity field of an
isolated disk galaxy, mainly in the northern part.
This field has an elongated
shape and is symmetric in the east-west direction.
In the zone of the spiral arm of the primary
galaxy of the pair the radial
velocity profiles are slighty broader than
those in the disk of NGC 5278. For NGC 5279
the outer zones of the disk show the presence
of double profiles. This fact means that the region of
double velocity profiles is confined to the periphery of the
disk of NGC 5279. In the disk of 
NGC 5278 the radial velocity values are 
in the range 7400--7860 km s$^{-1}$. 
Inside the disk region of NGC 5279 the radial 
velocity values are lower (7550-7650 km s$^{-1}$) 
than in the disk of NGC 5278.
The radial velocity of a bridge region
next to the disk of NGC 5278 lies in the range
7550-7570 km s$^{-1}$. Kinematical data and the morphological
shape of the pair indicate that there is a transfer 
of material between the two galaxies and suggest that 
the sense of transference is from NGC 5278 to NGC 5279.

\section{Rotation Curves}

Following \citet{Fuentes-Carrera2002, Fuentes-Carrera2004}
we obtain the rotation curve of each galaxy. The rotation
curve was obtained from the velocity field considering
the points within a given angular sector from the major axis.
The main caution is to exclude those points too close to the
minor axis leading to a strong dispersion of the points of
the rotation curve. The rejection of those points guarantee
us the symmetry of both sides of the rotation curve.
We can see from the velocity fields that the
inner parts of these two galaxies are not strongly
perturbed by the interaction process,
at least up to the radius $\approx$\,6
kpc ($\approx$12$\arcsec$) for NGC 5278 and
$\approx$\,5 kpc ($\approx$10$\arcsec$) for NGC 5279. 
Thus we can determine the rotation curve of both galaxies
considering a region of the velocity field within a
sector of a specified angle inside these radii.
\begin{figure*}[!htp]
\epsscale{1.0} \plotone{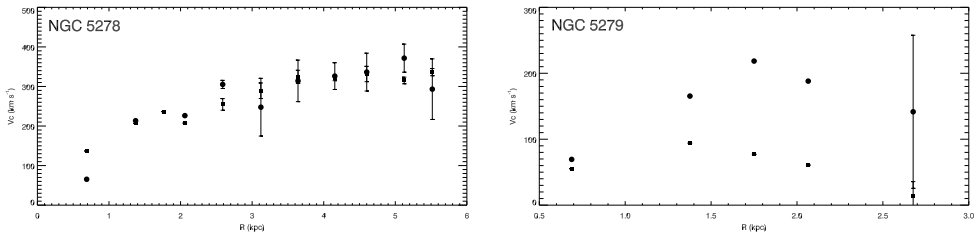} \caption{Rotation curves
of NGC 5278 and NGC 5279. Symbols correspond 
to approaching (filled circles) and receding (filled squares) 
sides. The vertical bars represent the dispersions of the different 
velocity points within the considered sector.\label{fig2}}
\end{figure*}
\section{NGC 5278/79: two trailing spirals}

\begin{figure*}[!htp]
\epsscale{1.0} \plotone{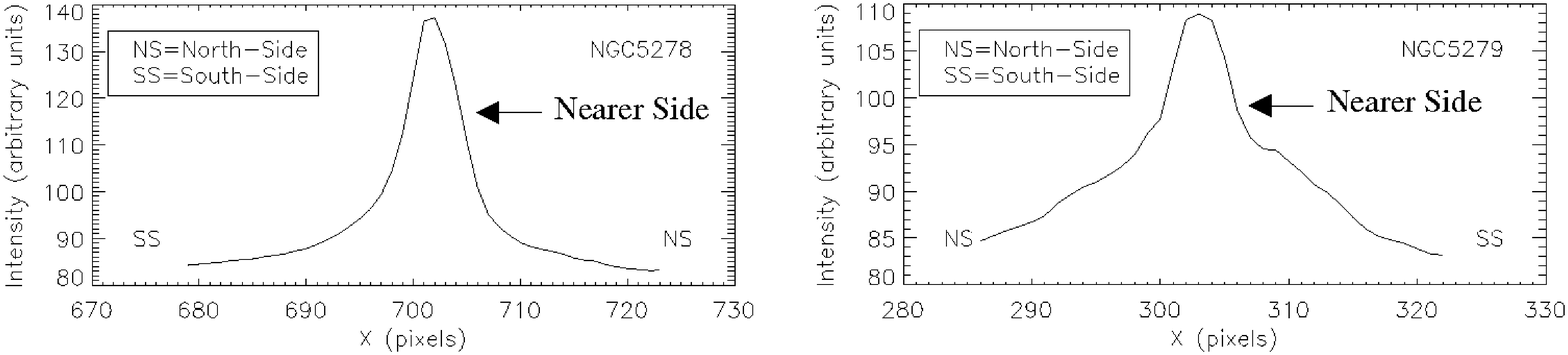}
\caption{Profile along the
kinematic minor axis of NGC 5278 and NGC 5279. From the top
profile it is clear that the northern--side is the nearer
in NGC 5278, because the profiles fall more abruptly
than along the southern--side. In the case of NGC 5279
the southern--side is the nearer because the profiles
fall more abruptly than along the northern--side, as
one can see from the bottom profile.\label{fig3}}
\end{figure*}

\begin{figure*}[!htp]
\epsscale{1.0} \plotone{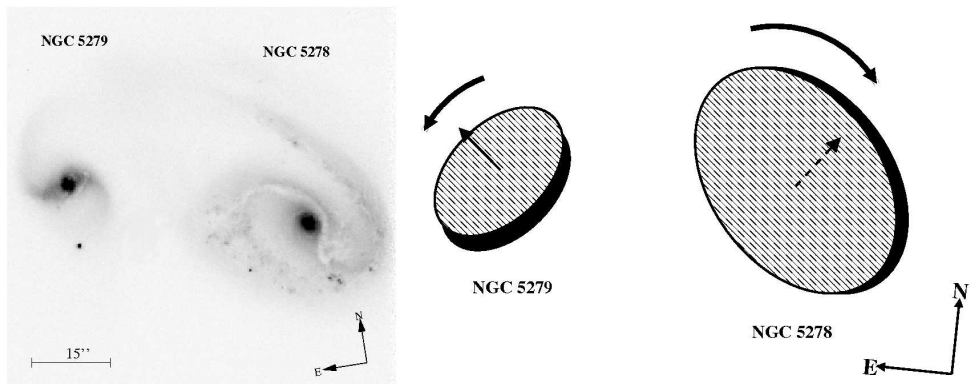}
\caption{HST image at $\lambda=$ 8230 \AA\, with the F814W filter 
\citep{Windhorst2002} toghether with a schematic picture of the 
interaction process for KPG 390.\label{fig4}}
\end{figure*}

This kinematic study sheds light on the
geometry of the galaxy encounter by determining 
the real orientation in the sky of the galaxy members as well
as the kind of spiral arms they possess. This later point is
not irrelevant in the case of interacting systems where a
possibility of having leading spiral arms is open.
Following \citet{Sharp1985} there is a criterion
that determines if any particular spiral galaxy has trailing
or leading arms. This criterion is based on three main
clues (receding-approaching side, direction of spiral arms
and the tilt of the galaxy, i. e., which side is closer to observer).
In our particular case, we have both the kinematic information
in order to establish which side of the galaxy is receding and
which side approaching, as well as very conspicuous morphological
aspects such as well defined spiral arm patterns and the presence
of dust lanes in both galaxy members running near the galaxy nuclei.
We use this last issue by getting an
intensity profile of the galactic nucleus along the minor axis.
In this kind of profiles, the nearest side is the steepest one
(because of the presence of the dust lane)\citep{Vaisanen2008}.
In the case of NGC 5278,
the receding radial velocities are in the south-western part, while the
approaching radial velocities are at the north-eastern side.
From Fig.~\ref{fig4} it is clear that the arms of
NGC 5278 are clockwise and the dust lane is
located at the concave side of the bulge, thus the northern side is
the nearest. This fact is confirmed by the profile extracted
along the kinematic minor axis of NGC 5278
(see Fig.~\ref{fig3}).
From these figures and the above criteria we have decided that NGC 5278 is
a trailing spiral because the sense of rotation is opposite to
the direction of the arms. We were able to apply similar
arguments to NGC 5279. In this case the receding radial
velocities are at the north-eastern side
of the galaxy and the approaching radial velocities are at
the south-western part. The arms in NGC 5279 are clockwise and
the nearest side is the southern side. As in the case of
NGC 5278 this fact is confirmed by the profile extracted
along the kinematic minor axis of NGC 5279 
(see Fig.~\ref{fig3}). We can conclude that
NGC 5279 is a trailing spiral also because the sense of rotation
is opposite to the direction of the arms.
A schematic representation of the interaction process is shown
in Fig.~\ref{fig4}.
\section{Conclusions}

In this article we presented Fabry-Perot observations
of the isolated pair of galaxies NGC 5278/79
(Arp 239, KPG 390) showing that for an interacting
and asymmetric system it is important
to have kinematic information of the entire field of
the galaxies participating in the interaction process.
We will use the kinematic information as a starting point
to fit the dark matter component and for preparing future numerical 
simulations of this pair.

\acknowledgements We acknowledge DGAPA-UNAM grant: IN102309
and CONACYT grant: 40095-F.

\end{document}